\documentclass[preprint,5p,times]{elsarticle}
\makeatletter
\def\ps@pprintTitle{%
 \let\@oddhead\@empty
 \let\@evenhead\@empty
 \def\@oddfoot{\textit{Preprint}\hfill\textit{2023}}%
 \let\@evenfoot\@oddfoot}
\makeatother

\usepackage{multirow}
\usepackage{hyperref}
\usepackage{dblfloatfix}
\usepackage[utf8]{inputenc}
\usepackage{outlines}
\usepackage{amssymb}
\usepackage{amsmath}
\usepackage{upgreek}
\usepackage{algorithm}
\usepackage{array}
\usepackage{bbold}

\usepackage{graphicx}
\usepackage{caption}
\usepackage{subcaption}
\usepackage[usenames]{color}
\usepackage[labelfont=bf,textfont=normalsize]{caption}
\usepackage[labelfont=bf,textfont=small]{subcaption}

\usepackage{here}
\usepackage{multicol}
\usepackage{wrapfig}
\usepackage{comment}
\usepackage{booktabs}
\usepackage{listings}
\usepackage{todonotes}
\usepackage{ulem}

\usepackage{verbatim}

\newcommand{\Delete} [1]{\bgroup\noindent\textcolor{red}{\xout{#1}}\egroup\ignorespacesafterend}

\usepackage{xcolor}

\date{2023}
\begin{document}

\begin{frontmatter}
\title{A graph database for feature characterization of dislocation networks}
\cortext[cor1]{Corresponding author.}

\author[IAM,HS]{Balduin Katzer}
\author[IPD]{Daniel Betsche}
\author[IPD]{Klemens Böhm}
\author[IAM]{Daniel Weygand}
\author[IAM,HS]{Katrin Schulz\corref{cor1}}

\address[IAM]{Karlsruhe Institute of Technology (KIT), Institute for Applied Materials (IAM),\\
              Kaiserstr. 12, 
              76131 Karlsruhe, Germany}
\address[HS]{Karlsruhe University of Applied Sciences, Moltkestr. 30, 76133, Karlsruhe, Germany}
\address[IPD]{Karlsruhe Institute of Technology (KIT), Institute for Program Structures and Data Organization (IPD),\\
              Am Fasanengarten 5, 
              76131 Karlsruhe, Germany}
\ead{katrin.schulz@kit.edu}
          
\begin{abstract}
Three-dimensional dislocation networks control the mechanical properties such as strain hardening of crystals.
Due to the complexity of dislocation networks and their temporal evolution, analysis tools are needed that fully resolve the dynamic processes of the intrinsic dislocation graph structure.
We propose the use of a graph database for the analysis of three-dimensional dislocation networks obtained from discrete dislocation dynamics simulations.
This makes it possible to extract (sub-)graphs and their features with relative ease.
That allows for a more holistic view of the evolution of dislocation networks and for the extraction of homogenized graph features to be incorporated into continuum formulation.
As an illustration, we describe the static and dynamic analysis of spatio-temporal dislocation graphs as well as graph feature analysis.
\end{abstract}
\end{frontmatter}

%
In dislocation theory, it is fundamental to understand how dislocations evolve, multiply or stabilize due to mutual interaction during plastic straining.
At mesoscopic length scales, dislocations are strongly interconnected with each other. 
This leads to the formation of complex three-dimensional dislocation networks.
The evolution behaviour of dislocation networks is fundamental for the resulting material properties~\cite{Nye1953,Hirth1982,KuhlmannWilsdorf1985}.
Dislocation networks and networks in general possess a graph structure~\cite{Estrada2012}.
Thus, one can subject it to many graph algorithms and graph frameworks~\cite{Po2014,ElAchkar2019,Starkey2022}.

In this work, we distinguish between the three-dimensional dislocation network (spatial topology) and its transformed graph theory based representation (graph topology).
For the transformation, the pristine three-dimensional data is imported into a graph database retaining all information of the original data, as described later.
This yields in the lower-level graph representation a so-called ''Property Graph''~\cite{Angles2018}.
The graph database allows us to perform graph analysis and to identify graph structures as well as to extract higher-level graphs, so-called hypergraphs~\cite{Dai2023}.
The graph topology is not static since the plastic deformation of a dislocation network incorporates the generation and dissolution of dislocations.
During this dynamic behavior the graph topology changes continuously. 
Thus, one can classify it as a spatio-temporal graph~\cite{DelMondo2010} with dynamic topology.

Previous studies on the evolution of three-dimensional dislocation networks within discrete dislocation dynamics (DDD) simulations have shown the richness of characteristics of these networks~\cite{Madec2008,Akhondzadeh2021, Katzer2022, Katzer2023}.
Madec et al.~\cite{Madec2008} observed specific dislocation reactions and related them to mechanical properties.
First approaches to graph analysis for dislocation networks already characterised defects under cyclic loading~\cite{ElAchkar2019}.
However, it has been shown that continuum models either have limited agreement with the data or do not include all details of complex network structures.
For example, extended continuum theories of dislocation reaction kinetics do only partially fit to three-dimensional DDD data~\cite{Katzer2022} or the analysis of complex dislocation mechanisms is only partially conducted for strongly interconnected structures~\cite{Akhondzadeh2021}.
The reason why these complex structures are often left unused is the difficulty in including them in the analysis.
Starting with the whole DDD simulation, we record it as a sequence of consecutive system states. 
Each state of this evolving system is stored as a snapshot of a graph at a discrete point in time.
A common technique of comparing two consecutive system states at times $t_{n}$ and $t_{n+1}$ is performing the comparison of the three-dimensional representation visually~\cite{Motz2009} or analysing each individual snapshot statistically~\cite{Katzer2023}.
Using only individual snapshots does not give way to any quantitative tracking of dislocations. 
By using a representation as spatio-temporal graphs, we aim at traceability of the dislocation network in space and time in quantitative terms.
A similar approach of temporal tracking has been introduced by Bertin et al.~\cite{Bertin2022}.
It converts graph snapshots to a continuum representation by Nye's tensor~\cite{Nye1953} in order to obtain an ''iso-topology''. 
This approach avoids the challenges of handling dynamic topology.
\begin{figure*}[ht]
    \centering
    \begin{subfigure}{\textwidth}
        \centering
        \includegraphics[width=\textwidth]{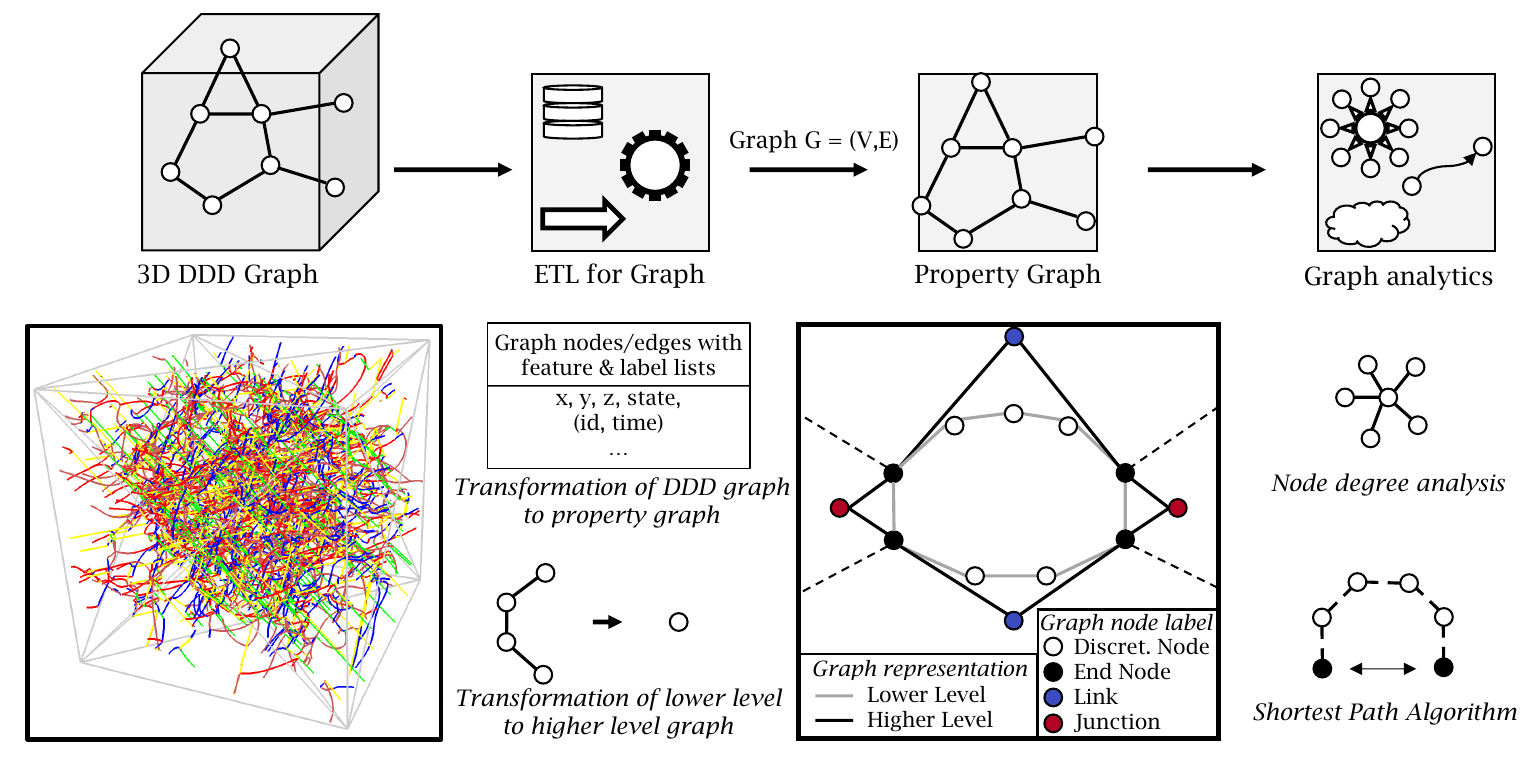}
    \end{subfigure}
   \caption{
   Data pipeline to the graph database: 
   Starting from a dislocation network of a three-dimensional DDD simulation, the spatial topology data is transformed into a property graph representation via the process of extracting, transforming, and loading (ETL).
   The property graph topology consists of nodes and edges, each of which has a list of features and labels that contain, for instance, spatial information or temporal information. This leads to a spatio-temporal graph, in which each object is defined by a tuple of identifier and time. An example of the property graph topology is illustrated by a lower-level graph representation, which represents the original discretized DDD data, and a higher-level graph representation, which is an aggregated graph consisting of dislocation links and junctions.
   Based on the graph representation, various graph analysis methods are applied such as shortest path algorithms or node degree analysis.
   }
   \label{fig:sketch}
\end{figure*}

In this work, we present a graph database containing the graph representation of converted three-dimensional dislocation network structures while preserving temporal and topological information.
Using a graph query language (GQL), we formulate our structures of interest as graph patterns with desired features, letting our database find the matches within our graph.
Additionally, we can employ graph algorithms to detect interesting dislocation constellations, e.g., shortest paths.
The result is a set of subgraphs that matches our graph patterns and feature values.
Using a graph representation and performing graph analysis is a powerful tool in other disciplines, e.g., social networks~\cite{Bedi2016}, traffic forecasting~\cite{Jiang2022} or drug discovery~\cite{Hall2017}.
The application of graph database technology to dislocation-based plasticity to perform analyses of complex but repeating structures with little effort, is the objective of this paper. 
Specifically we want to answer two question for metal plasticity in this paper: (i) ''How do dislocation graphs evolve in time and space'' and (ii) ''How do graph features influence the dynamic graph topology''?
We present our answers by facilitating static and dynamic analysis of dislocation graphs as well as extraction of graph features within our graph database.

We propose a graph database for the analysis of three-dimensional DDD simulations.
We use DDD according to~\cite{Weygand2001, Weygand2002} to conduct simulations of $5 \times 5 \times 5 \,\mu\textrm{m}^3$ tensile-tested face-centered cubic (fcc) single crystals mimicking aluminum.
For a detailed description of the simulation parameter and procedure, see~\cite{Katzer2022}.
Exemplarily for the DDD simulations, we use a high symmetrical $\langle100\rangle$ crystal orientation with a strain rate $\dot{\varepsilon}$ of $2000 \textrm{s}^{-1}$.
The initial relaxed dislocation network has a dislocation density $\rho_{0}$ of $1.2 \times 10^{13} \textrm{m}^{-2}$.

We use the graph database management system Neo4J, which is well known as a powerful open-source graph database~\cite{Angles2012}.
The process of extracting, transforming, and loading (ETL) our DDD simulation data from its source into our database follows a commonly established procedure~\cite{Kimball2004}.
This process is depicted schematically in~\autoref{fig:sketch}, where the spatial topology of the DDD data is transformed into a graph topology in the graph database.
The addition of features and labels to every edge and node turns our data structure into a ''Property Graph''~\cite{Angles2018}.
Formally, a graph $G = (V,E)$ is a collection of nodes, also called vertices, $v\in V$, and edges $e\in E$.
By adding a time value to turn each node $v$ into a tuple $v = (id, t)$ of its id and time, one obtains temporal graphs.
Analogously, we represent each edge $e$ as tuple $e = (id, t)$.
A snapshot is the subset of all nodes and edges in $G$ from the same observed state, i.e., the same point in time $t$. 
\begin{figure*}[ht]
\begin{minipage}[c][8.8cm][b]{.55\textwidth}
  \centering
  \includegraphics[width=\textwidth]{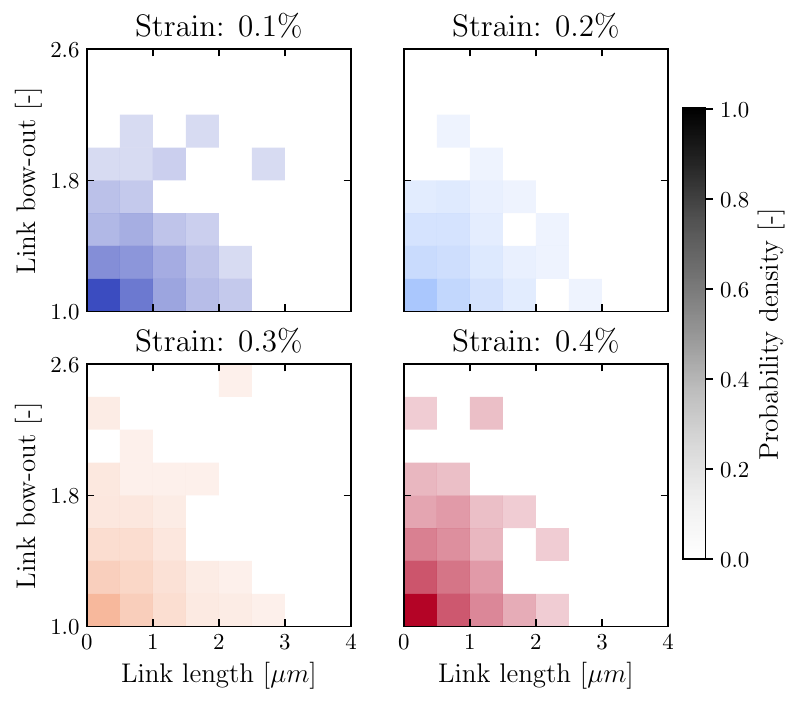}
\end{minipage}
\hspace{0.1cm}
\begin{minipage}[c][8.8cm][b]{.43\textwidth}
  \centering
  \includegraphics[width=\textwidth]{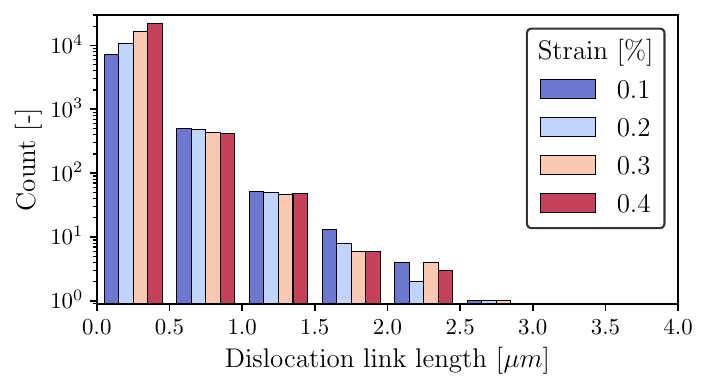}
  \includegraphics[width=\textwidth]{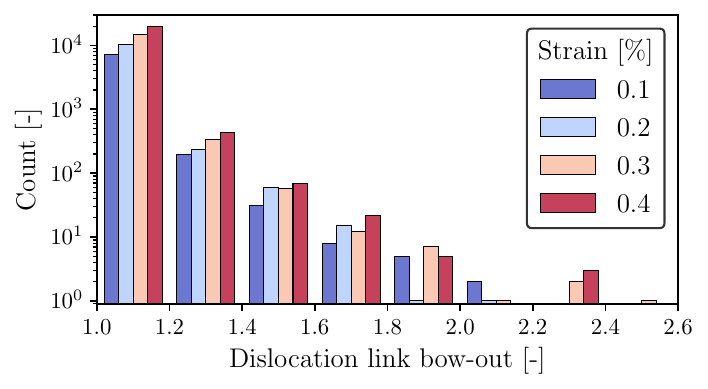}
\end{minipage}
\caption{Dislocation link bow-out vs link length at four distinct strain states (left). Distribution of dislocation link length and bow-out at these distinct strain states (right). For better visual comparability, the color of the strain states is transferred from (right) to (left) instead of a greyscale.}
\label{fig:static}
\end{figure*}

The transformation of the DDD data is conducted as follows: The three-dimensional DDD data is the spatial topology and consists of discretization nodes and their connection by edges, which represent dislocations.
The discretized node data is imported into the graph database, where spatial information such as, e.g., slip system and Burgers vector are stored as features in graph nodes resulting in a graph topology.
Similarly, the discretized edge data is imported as graph edges in the graph database connecting the graph nodes.
This results in a lower-level graph representation.
Within the graph database, we implement additional sets of higher-level graph nodes as e.g. ''junctions'' and ''dislocation links'', which are aggregates of all consecutively connected discretization nodes and edges of a same property, as introduced in~\cite{KuhlmannWilsdorf1985,Sills2018}.
Consequently, we add the label ''end node'' to a lower-level graph node if a dislocation link or a junction starts or ends in this node.
This label connects lower-level and higher-level graph representations.
The condensed information of links and junctions is stored as feature in the dislocation link, e.g., the line length or bow-out.
The bow-out is defined as the link length divided by the Euclidean distance of the two end nodes.
Ultimately, our graph database consists of a lower-level graph representation consisting of the pristine DDD data and a higher-level graph representation consisting of end nodes, links and junctions.
It should be remarked, that this procedure also allows for an easy implementation of even higher levels in the graph database.

For the temporal tracking of the dislocation graph, the creation of the $id$s is described briefly:
The link id is generated by its neighboring junction ids, i.e., the link id remains the same as long as the junction neighbor or the link itself does not interact. 
The junction id is created by its connected dislocation loop ids, which are already used for multiplication cascade tracking~\cite{Stricker2018} and are stored as a feature in the lower-level graph nodes.

In computer-science terminology, the dislocation network is a spatio-temporal graph.
This gives way to two options for analysis:
(i) The static analysis extracts information for each snapshot of the graph individually.
(ii) The dynamic analysis extracts information over all snapshots of the graph.
Going beyond mere statistical analysis, we make use of the GQL to match graph patterns.
Our database is an instance of Neo4j. 
Consequently, we formulate our queries in the GQL Cypher. 
This allows for efficient retrieval of graph data using a concise syntax.
To demonstrate the prospects of a graph database coupled with a query language for dislocation graphs, we present results of the static and dynamic analysis as well as some graph feature analyses of the dynamic topology.

The static analysis of the dislocation graph extracts statistics of each individual simulation state over time.
\autoref{fig:static} shows results of the static analysis of two graph features, dislocation link length and bow-out.
Considering the individual states, the link evolution shows a decrease in length and an increase in bow-out while straining the single crystal.
This observation holds true for almost the entire spectrum of lengths of dislocation links~(\autoref{fig:static}).
With this static analysis, we can query for information of each dislocation graph snapshot.

The dynamic analysis allows for a different view on the evolution of the graph over time compared to the static analysis.
It enables the tracing of the evolution of (sub-)graphs.
Based on the dynamic process of generation and dissolution of graph (sub-)structures, the dynamic analysis incorporates graph stability or instability analysis.
A (sub-)graph is considered stable as long as its graph topology does not change within at least one time step.
For example, physically, a stable graph topology may indicate that there is little or no plastic deformation, while an unstable graph may indicate increased plastic deformation.
Dislocation links undergo different length and bow-out changes while straining, as shown in~\autoref{fig:static}.
Besides dislocation generation, dislocation links can dissolve due to dislocation motion.
\begin{figure}[ht]
    \centering
    \begin{subfigure}{0.47\textwidth}
        \centering
        \includegraphics[height=7.5cm]{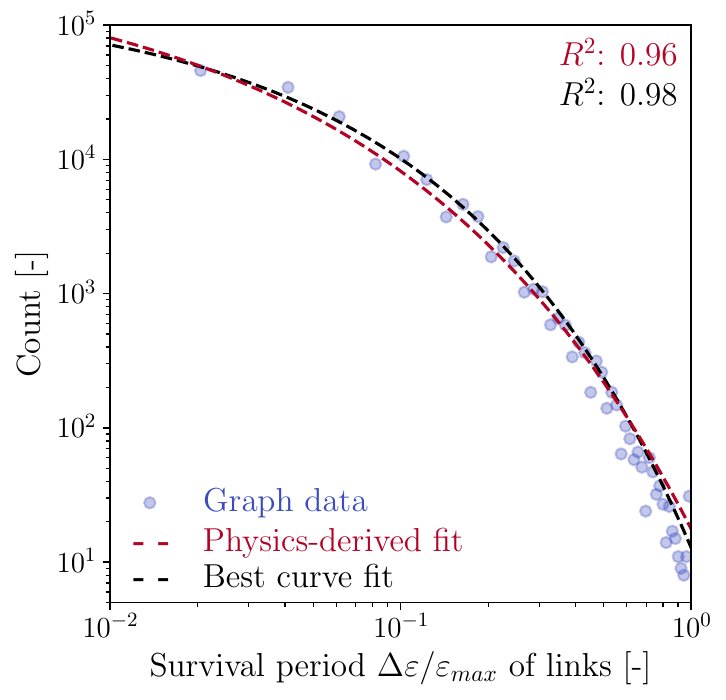}
    \end{subfigure}
   \caption{Survival period of dislocation links, with a best curve fit (black) and a physics-derived explanation (red).
   The physical derivation is based on a stretched exponential function of the form  $A \cdot exp(-(\tau_{d}^{-1} x)^{\beta})$, where the parameters $A$, $\tau_{d}^{-1}$ and $\beta$ are derived from the dislocation simulation.
   }
   \label{fig:dynamic}
\end{figure}

\autoref{fig:dynamic} shows the dislocation lifetime, i.e., the ''survival'' strain $\Delta\varepsilon$ of dislocation links with respect to the total strain $\varepsilon_{max} = 0.5\%$ of the simulation resulting in a survival period. 
The dislocation links either exist at the beginning of the simulation or are generated while straining.
We observe that the survival period of dislocation links tend to be short.
The survival period fits a stretched exponential function of the form $A \cdot exp(-(\tau_{d}^{-1} x)^{\beta})$.
We choose parameters physically as follows:
$A$ is the initial total number of links $x(0)$, $\tau_{d}^{-1}$ corresponds to a characteristic decay time $(\dot{\varepsilon}\cdot\Delta t)^{-1}$ and $\beta = 3/7$ is an exponent derived from considering short and long-range contributions~\cite{Macdonald2005,Mauro2012}.
Additionally, we add a best curve fit function.
Both functions are evaluated with the graph data by the coefficient of determination $\textrm{R}^2$, which is greater than $0.95$ in both cases.

\begin{figure*}[ht]
    \centering
    \begin{subfigure}{\textwidth}
        \centering
        \includegraphics[width=0.9\textwidth]{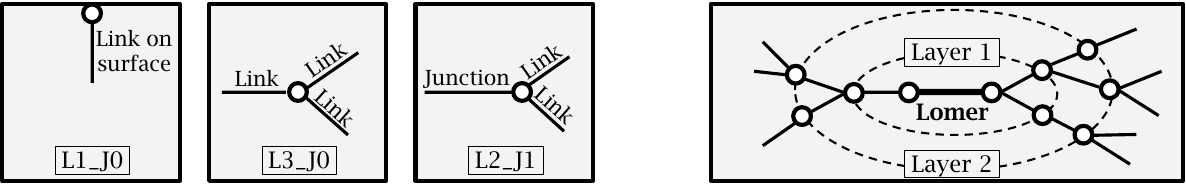}
    \end{subfigure}
    \begin{subfigure}{0.49\textwidth}
        \centering
        \includegraphics[height=7cm,
  keepaspectratio]{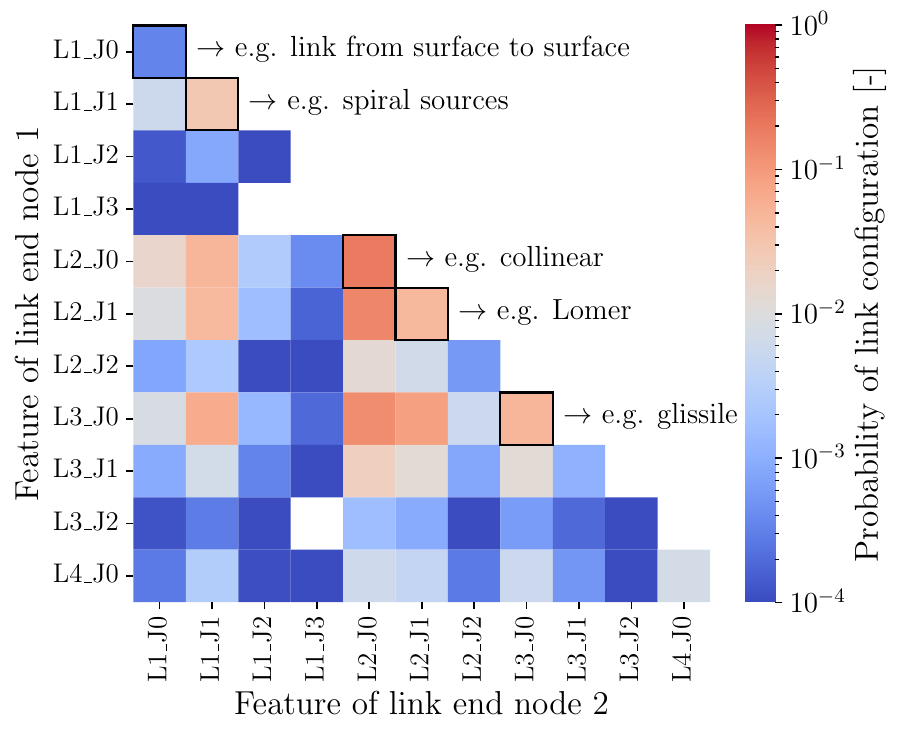}
    \end{subfigure}
    \begin{subfigure}{0.49\textwidth}
        \centering
        \includegraphics[height=7cm,
  keepaspectratio]{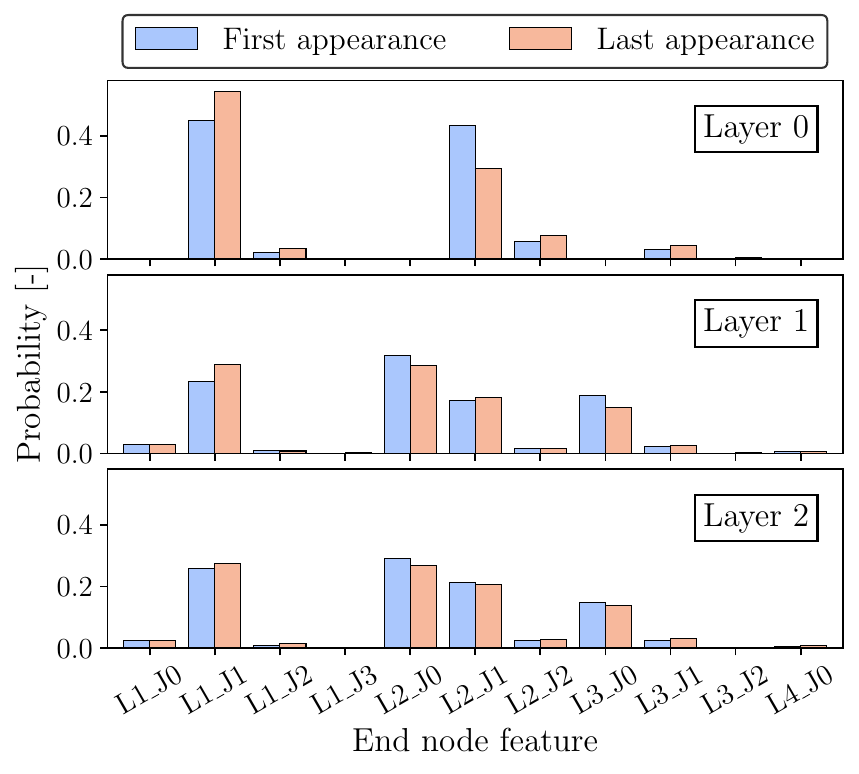}
    \end{subfigure}
   \caption{
   Probability of a link configuration based on its two end node features (left).
   The abbreviation of each end node feature ''LX\_JY'' represents the number of connected links to the node by X and the number of connected physical junctions (Lomer, Hirth) to the node by Y.
   Other node degrees exist, but are less than $0.1\%$.
   Link configurations are left blank if they are not observed.
   Examples of symmetric dislocation link configurations are highlighted by a frame.
   Probability of Lomer junction (layer 0) end node features and the probability of a Lomer junction's first (layer 1) and second (layer 2) nearest neighbor end node features at the first and the last appearance of a Lomer junction (right).
   }
   \label{fig:graph_feature}
\end{figure*}

A key concept in graph theory is the analysis of node degree, which refers to the number of edges connected to a node.
In the context of dislocation graphs, the degree is used to quantify the connectivity of end nodes to links or junctions to identify important nodes in the graph. 
End nodes with different degrees can lead to varying graph behavior.
An end node has X connected dislocation links and can have Y junctions (abbreviated by LX\_JY). 
For example, a simple Lomer reaction leads to two end nodes of type L2\_J1; a glissile reaction would lead to two L3\_J0 end nodes.
Thus, we define a set of end node features based on the number of connected links and junctions.

We analyze the two end node features of each dislocation link, see~\autoref{fig:graph_feature}(left).
We show the probability of link configurations, i.e. the probability that a link starts with one feature and ends with the same or a different feature.
The diagonal entries can correspond to configurations between the same junction type, e.g. a link from collinear to collinear (L2\_J0), from Lomer to Lomer (L2\_J1) or from glissile to glissle (L3\_J0), but can be of different junction type as well.
Additionally, there are surface to surface links (L1\_J0) or single-arm spiral sources (L1\_J1) between two Lomer junctions as shown in Motz et al.~\cite{Motz2009}.
However, we observe many configurations, where the two end nodes of a link consist of different features (off-diagonal entries).
The matrix is symmetric, since the end nodes of a link can be starting or end point in our analysis. 
One frequent off-diagonal example is a link, starting at a L3\_J0 (e.g. glissile) end node and ending in a L2\_J0 (e.g. collinear) or vice versa. 
Thus, the end node feature based analysis reveals 
that links of different end node features are frequent, which may influence the mobility.

Besides characterizing end nodes of link configurations, we use the graph topology to analyse the evolution of larger graph structures.
One interesting configuration are Lomer junctions and their k-nearest neighbors, since Lomer junctions are energetically favorable, but can be dissolved after a certain survival period.
We investigate, if the distribution of the surrounding end node features of a Lomer junction has an influence on its stability.
Therefore, we analyse Lomer junction configurations of the k-nearest neighbors at their first (one time step after creation) and their last (one time step before dissolution) appearance during the simulation during plastic deformation and with a survival strain of at least $0.2\%$.
\autoref{fig:graph_feature}(right) shows the probability of the Lomer junction (layer 0) end node features at its first and its last appearance, as well as the probability of the features of its first (layer 1) and its second (layer 2) neighboring end nodes.
Layer 0 reveals an decreased probability of L2\_J1 and an increased probability of L1\_J1 for the Lomer junction's last compared to its first appearance.
Thus, Lomer arms tend to react with ongoing plastic deformation without unzipping the Lomer junction, which is indicated by the end nodes features changing from L2\_J1 to L1\_J1.
Layer 1 shows that Lomer arms often end in L2\_J0 (e.g. collinear junction), in  L1\_J1 (Lomer/Hirth junction), or in L3\_J0 (e.g. glissile junction) end nodes.
We observe a slightly increased probability of L1\_J1 and a slightly decrease prdobability of L2\_J0 and L3\_J0 for the last compared to the first appearance.
The probability of end node features in layer 2 does not indicate an influence on the Lomer junction dissolution.
However, layer 0 and layer 1 show larger probability differences between first and last appearance of a Lomer junction.

The presented results are examples of query results that are easy to obtain once the dislocation network is modeled in a graph database.
The results demonstrate that a graph database is a promising tool for static and especially for dynamic analysis of dislocation graphs. 
The inherent mapping of the spatio-temporal features of dislocation networks to a graph representation leads to new insights into the evolution of dislocation networks.
Compared to the static analysis of Lomer arms~\cite{Katzer2023}, the graph database complements the spatial information with the temporal dimension by graph features.
The temporal tracking enables the analysis of generation and dissolution processes of mobile dislocation links by the survival strain analysis~(\autoref{fig:dynamic}).
Extending the static analysis, which deals with each snapshot graph individually~(\autoref{fig:static}), the full history of the temporal graph evolution is preserved in the dynamic graph analysis.
The stretched exponential function fit of the lifetime of dislocation links indicates strong changes of the graph topology over time~(\autoref{fig:dynamic}). 

The analysis of the link end node degrees yields interesting insights into the dislocation graph.
In contrast to simple link configuration concepts, the results shown in~\autoref{fig:graph_feature}(left) reveal a more complex picture including end node degrees larger than three and links starting and ending in different end node degrees.
This motivates the deployment of graph databases for even more detailed analyses of the structure of dislocation networks.
Specific structures like multi-junctions or second-order junctions have already been reported several times~\cite{Bulatov2006,Madec2008,Akhondzadeh2021}.
However, a systematic approach describing complex three-dimensional structures has been missing.
Our results show that a variety of other link and junction structures based on characteristics of the end node exist. 
We show that at creation and just before dissolution of a Lomer junction, the distribution of end node features directly at the Lomer junction changes as well as the distribution of end node features of its first and second nearest neighbors (\autoref{fig:graph_feature}(right)).
The analysis indicates that the neighboring structure seems to converge after two layers for a Lomer junction and could be seen as a characteristic structure.
The importance of the analysis of node degrees higher than three has already been discussed but only analysed to a limited extent , as e.g. for the so-called ''assisted glissile mechanism'' as one possible mechanism~\cite{Akhondzadeh2021}.
Therefore, we assume that end node degree analysis can reveal various other complex interaction mechanisms incorporating high node degrees.
Future research should demonstrate the derivation of flow rules incorporating dislocation network information such as link length (as in~\cite{Akhondzadeh2020}) or node degree.

Ultimately, the deployment of graph database technology should pave the way to study the inherent dynamic processes.
Further algorithms for temporal graphs, like minimal contrast subgraph pattern~\cite{Ting2006}, can help us in understanding the dislocation network evolution.
For example, comparing the subgraph of the last appearance of a Lomer junction and its k-nearest neighbors (\autoref{fig:graph_feature}(right)) with the subgraph after the dissolution of the Lomer junction can yield insights into junction dissolution processes.
Graph machine learning can be used to predict whole graph states, ultimately, with the goal to surrogate modelling. Hereby, the graph representation is useful, since we can convolve the graph into more condensed higher-level graphs (hypergraphs) to reduce the size of the graph for a faster prediction.
Finally, we demonstrated the applicability of graph databases to analyse the evolution of dislocation networks, but this technology is applicable to any graph representation from materials science and engineering, which could include converted experimental data~\cite{Sills2022}. 
%

\section*{Acknowledgement}
\label{sec:Acknowledgement}
The  financial  support for this work in the context of the DFG research projects SCHU 3074/4-1 and BO 2129/16-1 is gratefully acknowledged.
This work was performed on the HoreKa supercomputer funded by the Ministry of Science, Research and the Arts Baden-Württemberg and by the Federal Ministry of Education and Research.
This work was supported by the Ministry of Science, Research and the Arts Baden Württemberg, project Algorithm Engineering for the Scalability
Challenge (AESC).
\section*{Declaration of Competing Interest}
\label{sec:Declaration}
The authors declare that they have no known competing financial interests or personal relationships that could have appeared to
influence the work reported in this paper.
\bibliographystyle{elsarticle-num}
\bibliography{GraphDatabase.bib}
\end{document}